\documentclass[aps,preprint, nofootinbib,floatfix,superscriptaddress]{revtex4}
\usepackage{float}
\usepackage[final]{graphicx}
\usepackage{epsfig}
\usepackage{epstopdf}
\usepackage{graphicx}
\usepackage{multirow}
\usepackage{amsmath,stackrel}
\usepackage{ulem}
\usepackage{bm}
\usepackage{footnote}
\usepackage{color}
\def\nn {\nonumber}
\newcommand{\be}{\begin{equation}}
\newcommand{\ee}{\end{equation}}
\newcommand{\bea}{\begin{eqnarray}}
\newcommand{\eea}{\end{eqnarray}}

\newcommand{\bx}{\bm x}

\newcommand{\br}{\bm r}
\newcommand{\bk}{\bm k}
\newcommand{\bp}{\bm p}

\def\llcol#1#2{\tilde{\lambda}_{#1}.\tilde{\lambda}_{#2}}
\begin{document}
\title{Open-Flavor Heavy Hadron Production in Heavy-Ion Collisions}
\author{C.~E.~Fontoura}
\email{ce.fontoura@unesp.br}
\affiliation{Centro Universit\'ario Faculdade de Inform\'atica e Administraç\~ao Paulista, S\~ao Paulo, SP, Brazil} 
\affiliation{Instituto de F\'{\i}sica Te\'orica, Universidade Estadual Paulista,
Rua Dr. Bento Teobaldo Ferraz, 271 - Bloco II, 01140-070, 
S\~ao Paulo, SP, Brazil}
\author{G.~Krein}
\email{gastao.krein@unesp.br}
\affiliation{Instituto de F\'{\i}sica Te\'orica, Universidade Estadual Paulista,
Rua Dr. Bento Teobaldo Ferraz, 271 - Bloco II, 01140-070, 
S\~ao Paulo, SP, Brazil}

\author{A.~Valcarce}
\email{valcarce@usal.es}
\affiliation{Departamento de F\'\i sica Fundamental and IUFFyM,
Universidad de Salamanca, 37008 Salamanca, Spain}
\author{J.~Vijande}
\email{javier.vijande@uv.es}
\affiliation{Unidad Mixta de Investigacin en Radiof\'{\i}sica e Instrumentaci\'on Nuclear en Medicina (IRIMED),
Instituto de Investigaci\'on Sanitaria La Fe (IIS-La Fe) \\
Universitat de Valencia (UV) and IFIC (UV-CSIC), Valencia, Spain}

\begin{abstract}
We study the production of open-flavor heavy hadrons
in relativistic heavy-ion collisions. The hadronization in the quark-gluon plasma is described in the quark coalescence model. We evaluated yields and transverse momentum distributions. A simultaneous study of conventional and exotic hadrons is carried out. The Wigner functions are evaluated using hadron wave functions obtained from a single realistic quark model. Thus, results are presented in a single framework for the production of open-flavor heavy mesons, baryons, and exotic tetraquarks, in particular: $D^{0}$, $B^{0}$, $\Lambda_{Q}$, $\Sigma_{Q}$, $\Xi_{Q}$, $\Xi_{QQ^\prime}$, and $T_{QQ^\prime}$ ($Q,Q^\prime=c$ or $b$). The consequences of a partial restoration of chiral symmetry at the hadronization temperature are studied in detail.
\end{abstract}
\maketitle
\section{Introduction}
\label{sec:introduction}

There has been significant experimental progress in recent years in the spectroscopy of heavy hadrons~\cite{Nav24}, both in conventional mesons composed of a simple quark-antiquark pair and baryons composed of three valence quarks, as well as in those with unconventional quark content such as tetraquarks and pentaquarks, known as exotic hadrons.  A very ambitious program is being developed at the Large Hadron Collider (LHC), one of whose objectives is the discovery and study of new heavy hadrons either through proton-proton ($pp$) collisions (LHCb) or heavy-ion ($AA$) collisions (ALICE). The Relativistic Heavy Ion Collider (RHIC), the second-highest-energy heavy-ion collider in the world, is involved in similar studies. This type of experiment has already reported the observation of various new types of heavy-flavor hadrons, e.g., $X$~\cite{Aaij:2013}, $\Omega_c$~\cite{Aaij:2017nav}, $\Xi_{cc}^{++}$~\cite{Aaij:2017} and $T_{cc}$~\cite{{Aai21},{Aaj21}}. 

Relativistic heavy-ion collisions, while primarily aimed at studying the elusive state of matter known as quark-gluon plasma (QGP), also offer a wide range of applications. Among these, the search for new hadronic structures within the QGP represents an exciting new direction in the quest to deepen our understanding of heavy-flavor hadron spectroscopy. In the heavy-flavor sector, there is a broad consensus on the key role of the coalescence mechanism in the description of the hadronization process within the QGP~\cite{Cho:2017dcy}. 

Within this perspective, in the present work, we study the production of open-flavor heavy hadrons in relativistic heavy-ion collisions by coalescence hadronization. We employ the dynamical coalescence model, extensively used in hadron production~\cite{Cho:2017dcy}, to study the yields, ratios, and transverse momentum distributions of conventional and exotic open-flavor heavy hadrons. We will focus on hadrons made of one or two heavy quarks, in particular: $D^{0}$, $B^{0}$, $\Lambda_{Q}$, $\Sigma_{Q}$, $\Xi_{Q}$, $\Xi_{QQ^\prime}$, and $T_{QQ^\prime}$ ($Q, Q^\prime=c$ or $b$). $T_{QQ^\prime}$ stands for the $J^P=1^+$ $QQ^\prime \bar u\bar d$ tetraquark. A key ingredient of the coalescence model is the Wigner function, which describes the spatial and momentum distribution of quarks in a hadron and, therefore, depends critically on the hadron wave function. It is therefore very important that the structure of the hadrons studied be described by realistic models, explaining basic properties such as binding energies and root mean square radii. This is especially important for the production ratios between different types of hadrons, which are among the best studied observables at the LHC and RHIC. Thus, we calculate the Wigner functions employing the two-, three- and four-body wave functions obtained from a single realistic constituent quark model that correctly reproduces the low-lying meson and baryon spectra~\cite{Sem94} as well as the bound nature of the exotic $QQ\bar n\bar n$ tetraquarks~\cite{Ade82}.

This type of study is of relevance especially for the possible production of stable bottom tetraquarks. Since the discovery of  $T_{cc}$~\cite{{{Aai21},{Aaj21}}}, there is a general consensus on the nature of its bottom cousin with a larger binding energy~\cite{Col24}. Its discovery would represent a milestone of great importance for hadronic physics, as it is a prediction more than forty years old~\cite{Ade82}. Therefore, having realistic predictions of the production of such exotic hadrons compared to the known heavy hadrons is of paramount importance in the design of experiments searching for new states of matter.

Specifically, we study open-flavor heavy-hadron production at the energies available at LHC and RHIC, in particular in central ${\rm Au} + {\rm Au}$ collisions at RHIC at $\sqrt{s_{NN}} = 200$~GeV and in ${\rm Pb} + {\rm Pb}$ collisions at LHC at $\sqrt{s_{NN}} = 2.76$~TeV and $\sqrt{s_{NN}} = 5.02$~TeV. Hadrons are produced in the hot and dense environment of a relativistic heavy-ion collision. Putting it in another way, coalescence happens at nonzero temperature, at which the coalescing (light) constituent quarks have properties different from those in vacuum, and thus the hadron wave functions are expected to be modified. 
Therefore, we also address the important question of how partial restoration of chiral symmetry affects the coalescence process~\cite{Fries:2008hs}.

The paper is organized as follows. In Sect.~\ref{sec:1} we briefly review the basic features of the coalescence model. We discuss the structure of the hadron wave functions employed in this work and its use in the computation of the Wigner function. We present the variation of the hadron properties with temperature that will influence the evaluation of the Wigner function. In Sect.~\ref{sec:5} we present and discuss our results for the yields, ratios, and transverse momentum distributions of the different hadrons, in comparison with previous estimates in the literature and experimental data. We will show results for the most promising exotic tetraquark candidates, the isoscalar doubly heavy axial-vector tetraquarks $T_{QQ^\prime}$. Finally, in Sect.~\ref{sec:6} we summarize the most important findings of our work.

\section{Coalescence of quarks in hadrons}
\label{sec:1}

In the coalescence model, the probability of producing hadrons from quarks in the medium formed in a QGP is given by the overlap of the Wigner function of the produced hadron with the phase-space distribution of the constituents in the medium. In a covariant approach, the momentum spectrum of hadrons formed by the coalescence of quarks can be written as~\cite{{Greco:2003mm},{Greco:2003xt}}:
\be
\frac{d^2N_{h}}{d^2\bp_T}=\frac{g_{h}}{V^{n_q-1}}
\int\,\prod^{n_q}_{i=1}\left[\,d^2\bp_{Ti}\, \frac{d^2N_{q_i}}{d^2\bp_{Ti}}\,\right]\delta^{(2)}\left(\bp_T-\sum^{n_q}_{i=1}\bp_{Ti}\right) \rho_{h}\left(\bk_1,\cdots,\bk_{n-1}\right) \, , 
\label{3CoalTrans}
\ee
where $n_q$ is the number of constituent quarks, $V$ is the volume of the source, $g_h$ is a statistical factor, and $\rho_{h}$ is the Wigner function. 
The Wigner function describes the spatial and momentum distribution of quarks in a hadron and can be related to the hadron wave function as:
\bea
\rho_{h}(\bk_1, \cdots\bk_{n_q-1}) &=&
\int\,\left(\prod^{n_q-1}_{j=1}d^3\br_{j}\,d^3\br'_{j}\,e^{-i \bk_j\cdot r'_{j}}\,\right)
\psi\left(\br_1 + \frac{\br'_1}{2},\cdots, \br_{n_q-1} + \frac{\br'_{n_q-1}}{2}\right)
 \nn \\[0.1cm]&\times&
\psi^{*}\left(\br_1 - \frac{\br'_1}{2},\cdots, \br_{n_q-1} -\frac{\br'_{n_q-1}}{2}\right).
\label{wigner-jacobi}
\eea
\subsection{The Wigner function}
\label{sec:Wigner}
For $n_q=2$, Eq.~\eqref{3CoalTrans} describes the production
of a meson. The meson wave function can be written as
\be
\psi(\br) = \chi^{csf}\, R(\br) \, ,
\label{meson-trial-wave-function}
\ee
where $\chi^{csf}$ are orthonormalized color-spin-flavor vectors and $R(\br)$ is the radial part. It can be expanded in terms of Gaussians,
\be
R(\br)=\frac{1}{4\pi}\sum^{n_{g}}_{i=1}
\alpha_{i}\, e^{-a^{i}\,\br^{2}} \, ,
\label{meson_wf}
\ee
where $a^{i}=1/(2\,s^{2}_{i})$, $n_g$ is the number of Gaussians and $s_{i}$ are variational parameters determined by minimizing the mass of the hadron (meson) for a particular interquark potential. The integration of the spatial part of the Wigner function can be performed analytically and the result can be written as:
\be
\rho_{M}(\bk) = \frac{1}{4\pi}\sum^{n_{g}}_{i,j=1}\alpha_{i}\,\alpha_{j}\int\,d^{3}\br\,d^{3}\br'\,e^{-i\bk\cdot \br'}e^{-{\cal E}_W(\br^\pm)} \, ,
\label{wigner-jacobi-meson}
\ee
where $\br^\pm = \br\pm \br'/2$ and, 
\be
{\cal E}_W(\br^\pm) =
a^{i}\,(\br^{+})^2 + a^{j}\,(\br^{-})^2 \, .
\label{EWM}
\ee
The relative coordinate $\br$ and the transverse momentum $\bk$ between the quark and the antiquark are defined by,
\bea
\br &=&\bx_{1}-\bx_{2} \, , \nn
\\[0.1cm]
\bk &=& \frac{m_2\,\bp_1 - m_1\,\bp_2}{m_1 + m_2} \, .
\label{meson-rel-mom}
\eea

For baryons ($n_q=3$), the wave function is expressed as a sum over all allowed channels with well-defined symmetry properties, namely:
\be
\psi(\br_{1},\br_{2}) = \sum^{n}_{\kappa=1} \chi^{csf}_{\kappa} \,
R_{\kappa}(\br_{1},\br_{2}),
\label{baryon-trial-wave-function}
\ee
where $n$ is the number of channels, $\chi^{csf}_\kappa$ are orthonormalized color-spin-flavor vectors
and $R_{\kappa}(\br_{1},\br_{2})$ is the radial part of the wave function of the $\kappa-$th
channel. Each $ R_\kappa(\br_{1},\br_{2})$ is expanded in terms of generalized Gaussians,
\be
R_\kappa(\br_{1},\br_{2}) = \sum^{n_{g}}_{i=1} \alpha^{\kappa}_i \,
e^{-a^{i}_\kappa\,\br^{2}_{1} - b^{i}_\kappa \,\br^{2}_{2} - c^{i}_\kappa \,\br_{1}\cdot\br_{2}} \, ,
\label{Rkr}
\ee
where $a^{i}_\kappa,b^{i}_\kappa$, and $c^{i}_\kappa$ are variational parameters determined, as for mesons, by minimizing the mass of the baryons for some particular interacting potential between the quarks.
The Jacobi coordinates $\br_i$ and momenta $\bk_i$ of the baryon are defined by:
\bea
\br_{1}&=&\bx_{1}-\bx_{2} \, , \nn
\label{bar-rel-coor1}
\\[0.1cm]
\br_{2}&=&\frac{m_{1}\,\bx_{1} + m_{2}\,\bx_{2}}{m_{1} + m_{2}}-\bx_{3}\, , \nn
\label{bar-rel-coor2}
\\[0.1cm]
\bk_{1}&=&\frac{m_2\,\bp_1 - m_1\,\bp_2}{m_1 + m_2}\, , \nn
\label{bar-rel-mom1}
\\[0.1cm]
\bk_{2}&=&\frac{m_{3}\,(\bp_{1} +\,\bp_{2})-(m_{1}+m_{2})\bp_{3}}{m_{1} + m_{2}+m_{3}}\, .
\label{bar-rel-mom2}
\eea
As in the case of mesons, an analytical expression for the Wigner function can be obtained:
\be
\rho_{B}(\bk_1,\bk_2)
= \sum^{n}_{k=1}\sum^{n_{g}}_{i,j=1} \alpha^k_i \, \alpha^k_j
\int\,\left(\prod^2_{m=1} d\br_m\,d\br'_m \, e^{-i \bk_m\cdot \br'_m}\right)  \,e^{- {\cal E}_W(\br^\pm_1, \br^\pm_2)} \, , 
\label{wigner-jacobi-baryon}
\ee
where $ \br^\pm_i = \br_i \pm \br'_i/2$ and, 
\be
{\cal E}_W(\br^\pm_1, \br^\pm_2) =
a^{i}_k (\br^+_1)^2 
+ b^{i}_k (\br^+_2)^2
+ a^{j}_k (\br^-_1)^2
+ b^{j}_k (\br^-_2)^2
+ c^{i}_k \br^+_1\cdot \br^+_2
+ c^{j}_k \br^-_1\cdot \br^-_2
\label{EWB}\,.
\ee

For tetraquarks ($n_q = 4$), the wave function is expanded in terms of channels with well-defined symmetry properties, including hidden color channels~\cite{Vijande:2009ac,Caramees:2018oue}: 
\be
\psi(\br_{1},\br_{2},\br_{3}) = \sum^{n}_{\kappa=1} \chi^{csf}_{\kappa} \, 
R_{\kappa}(\br_{1},\br_{2},\br_{3}) \, ,
\label{trial-wave-function}
\ee
where $\chi^{csf}_\kappa$ are orthonormalized  color-spin-flavor vectors and $R_{\kappa}(\br_{1},\br_{2},\br_{3})$ is the radial part of the wave function of the $\kappa-$th channel. To guarantee the appropriate symmetry properties, $R_{\kappa}(\br_{1},\br_{2},\br_{3})$ is expressed as the sum of four components: 
\be
R_{\kappa}(\br_{1},\br_{2},\br_{3}) = \sum^4_{r=1} w(\kappa,r)  \, R^r_\kappa(\br_{1},\br_{2},\br_{3}) \, ,
\label{Rk}
\ee
where $w(k,r) = \pm 1$. Finally, each $ R^r_\kappa(\br_{1},\br_{2},\br_{3})$ is expanded in terms of $n_g$ generalized Gaussians as follows:
\be
R^r_\kappa(\br_{1},\br_{2},\br_{3}) = \sum^{n_g}_{i=1} \alpha^{\kappa}_i \,
e^{ - a^{i}_\kappa\,\br^{2}_{1} - b^{i}_\kappa \,\br^{2}_{2} - c^{i}_\kappa \,\br^{2}_{3}
- d^{i}_\kappa \, s_1(r) \,\br_{1}\cdot\br_{2} - e^{i}_\kappa \, s_2(r) \,\br_{1}\cdot\br_{3}
- f^{i}_\kappa \, s_3(r) \,\br_{2}\cdot\br_{3}} \, ,
\label{Rkr2}
\ee
where $s_1(r),\cdots,s_3(r)$ are equal to $\pm 1$ and $a^{i}_\kappa,\cdots, f^{i}_\kappa$ are variational parameters. Again, working with generalized Gaussians allows us to obtain an analytical expression for the Wigner function that can be written as:
\be
\rho_{T}(\bk_1,\bk_2,\bk_3)=
\sum^n_{\kappa=1}\sum^4_{r,r'=1}w(\kappa,r) w(\kappa,r')\sum^{n_g}_{i,j=1} \alpha^\kappa_i \alpha^\kappa_j
\int\,\left(\prod^3_{m=1}\,d\br_m \, d\br'_m \, e^{-i \bk_m\cdot \br'_m}\right)  \,e^{- {\cal E}_W(\br^\pm_1, \br^\pm_2, \br^\pm_3)} \, , 
\label{wigner_final_tetratquark}
\ee
where $ \br^\pm_i = \br_i \pm \br'_i/2$ and,
\bea
{\cal E}_W(\br^\pm_1, \br^\pm_2, \br^\pm_3) &=&
a^{i}_k (\br^+_1)^2 
+ b^{i}_k (\br^+_2)^2
+ c^{i}_k(\br^+_3)^2
+ a^{j}_k (\br^-_1)^2
+ b^{j}_k (\br^-_2)^2
+ c^{j}_k (\br^-_3)^2
\nn\\[0.2cm]
&&
+ d^{i}_k s_1(r) \br^+_1\cdot \br^+_2
+ e^{i}_k s_2(r) \br^+_1\cdot \br^+_3
+ f^{i}_k s_3(r) \br^+_2\cdot \br^+_3
+ d^{j}_k s_1(r) \br^-_1\cdot \br^-_2
\nn\\[0.2cm]
&&
+ e^{j}_k s_2(r) \br^-_1\cdot \br^-_3
+ f^{j}_k s_3(r) \br^-_2\cdot \br^-_3
\label{EW} \, .
\eea
In the above, we have used the following set of Jacobi coordinates:
\bea
\br_{1} &=& \bx_1 - \bx_2 \, , \nn
\label{tetra-rel-coord1}
\\[0.1cm]
\br_{2} &=& \bx_3 - \bx_4 \, , \nn
\label{tetra-rel-coord2}
\\[0.2cm]
\br_{3} &=& \frac{m_1\,\bx_1 + m_2 \,\bx_2}{m_1+m_2} -\frac{m_3\, \bx_3 + m_4 \,\bx_4}{m_3+m_4}\, ,
\label{tetra-rel-coord3}
\eea
and the corresponding momenta: 
\bea
\bk_1 &=&\frac{m_{2}\,p_{1} - m_{1}\,p_{2}}{m_{1}+m_{2}} \, , \nn
\label{tetra-rel-mom1}
\\[0.1cm]
\bk_2 &=&\frac{m_{4}\,p_{3} - m_{3}\,p_{4}}{m_{3}+m_{4}} \, , \nn
\label{tetra-rel-mom2}
\\[0.1cm]
\bk_3 &=&\frac{(m_{3}+m_{4})\,(p_{1}+p_{2}) - (m_{1}+m_{2})\,(p_{3}+p_{4})}{m_{1}+m_{2}+m_{3}+m_{4}} \, .
\label{tetra-rel-mom3}
\eea

\subsection{The hadron wave function and mass}
\label{sec:hamiltonian}
The hadron wave functions are obtained from the solution
of the Schr\"odinger equation, 
\be
H = \sum_{i} \left(m_{i}+\frac{p^{2}_{i}}{2m_{i}}\right)+\sum_{i<j}\,V(r_{ij}),
\label{QM-H}
\ee
where $V(r_{ij})$ is taken to be a realistic quark-quark potential.
We use a generalized Gaussian variational method,  
particularly well-suited for addressing the low-lying states of few-body
systems. See Ref.~\cite{Vijande:2009ac} for further details about the minimization procedure.

We adopt a generic constituent model, containing chromoelectric and 
chromomagnetic contributions, tuned to reproduce the masses of the mesons and baryons, the so-called AL1 model by Semay and Silvestre-Brac~\cite{Sem94}, widely used in 
a number of exploratory studies of multiquark systems~\cite{Sil93,Jan04,Her20,Hiy18}. It includes a standard 
Coulomb-plus-linear central potential, supplemented by a smeared version of the chromomagnetic interaction,
\begin{align}
\label{ecu3}
V(r)  & =  -\frac{3}{16}\, \llcol{i}{j}
\left[\lambda\, r - \frac{\kappa}{r}-\Lambda + \frac{V_{SS}(r)}{m_i \, m_j}  \, \vec \sigma_i \cdot \vec\sigma_j\right] \, ,\\ \nonumber \\
V_{SS}  &= \frac{2 \, \pi\, \kappa^\prime}{3\,\pi^{3/2}\, r_0^3} \,\exp\left(- \frac{r^2}{r_0^2}\right) ~,\quad
 r_0 =  A \left(\frac{2 m_i m_j}{m_i+m_j}\right)^{-B} \, , \nonumber
\end{align}
where
$\lambda=$ 0.1653 GeV$^2$, $\Lambda=$ 0.8321 GeV, $\kappa=$ 0.5069, $\kappa^\prime=$ 1.8609,
$A=$ 1.6553 GeV$^{B-1}$, $B=$ 0.2204, $m_u=m_d=$ 0.315 GeV, $m_s=$ 0.577 GeV, $m_c=$ 1.836 GeV and $m_b=$ 5.227 GeV. 
Here, $\llcol{i}{j}$ is a color factor, suitably modified for the quark-antiquark pairs.
Note that the smearing parameter of the spin-spin 
term is adapted to the masses involved in the quark-quark or quark-antiquark pairs. 
The parameters of the AL1 potential are constrained in a 
simultaneous fit of 36 well-established mesons and 53 baryons,
with a remarkable agreement with data, as could be seen in Table~2 of Ref.~\cite{Sem94}.
It is worth to note that although the $\chi^2$ obtained in Ref.~\cite{Sem94} with the AL1 potential
is slightly larger than the one obtained with other models, this is essentially
because a number of resonances with high angular momenta were considered. The AL1 model
is very well suited to study the low-energy hadron spectra~\cite{Sil96}. 

Finite temperature effects are incorporated in the parameters of the model that are related to the dynamical breaking of chiral symmetry, namely the masses of the light constituent quarks predicted by the Nambu-Jona-Lasinio model~\cite{Nam61} -- see Refs.~\cite{{Carames:2016qhr},{Fontoura2019}} for further details. It is important to note that the mass of the heavy quarks remains unaffected by changes in temperature.
In Table~\ref{Tab1} we present the temperature dependence of 
the masses of open-charm and bottom mesons, baryons and tetraquarks
predicted by the AL1 model discussed above. 
\begin{table}[t]
\caption{ Masses (in MeV) of light quarks and open-charm and bottom
mesons, baryons and tetraquarks for different 
temperatures (in MeV).}
\begin{center}
\begin{tabular}{lcccccc}
\hline\hline\\[-0.5cm]
&\multicolumn{1}{c}{Vacuum}&&\multicolumn{3}{c}{\hspace{2cm}In-medium}&\\
\cline{2-2}\cline{4-7}\\[-0.5cm]
                   & \hspace{-0.3cm} $T=0$     &   & $T=100$   & \hspace{0.3cm}$T=120$   & \hspace{0.3cm}$T=140$   & \hspace{0.1cm}$T=156$ \\
\hline\\[-0.6cm]
$m_{u,d}$            &$315$      &   &$304$    &$288$      &$256$     &$221$\\ \hline
$M_{D^{0}}$          & $1862$    &   & $1862$  & $1863$    & $1867$   & $1880$\\
$M_{B^{0}}$          & $5293$    &   & $5294$  & $5299$    & $5306$   & $5324$\\
\hline
$M_{\Lambda_{c}}$    & $2281$    &   & $2270$  & $2243$    & $2217$   & $2176$\\
$M_{\Sigma_{c}} $    & $2456$    &   & $2453$  & $2451$    & $2455$   & $2472$\\
$M_{\Xi_{c}}$        & $2464$    &   & $2461$  & $2456$    & $2453$   & $2455$\\
$M_{\Xi_{cc}}$       & $3607$    &   & $3607$  & $3607$    & $3611$   & $3623$\\
\hline
$M_{\Lambda_{b}}$    & $5632$    &   & $5623$  & $5599$    & $5578$   & $5545$\\
$M_{\Sigma_{b}}$     & $5844$    &   & $5843$  & $5846$    & $5853$   & $5878$\\
$M_{\Xi_{b}}$        & $5802$    &   & $5800$  & $5796$    & $5795$   & $5800$\\
$M_{\Xi_{bc}}$       & $6914$    &   & $6914$  & $6916$    & $6921$   & $6936$\\
$M_{\Xi_{bb}}$       & $10195$   &   & $10195$ & $10199$   & $10206$  & $10223$\\
\hline
$M_{T_{cc}}$         & $3877$    &   & $3871$  & $3857$    & $3845$   & $3829$\\
$M_{T_{bc}}$         & $7131$    &   & $7127$  & $7118$    & $7111$   & $7096$\\
$M_{T_{bb}}$         & $10410$   &   & $10402$ & $10395$   & $10378$  & $10356$\\
\hline\hline
\end{tabular}
\end{center}
\label{Tab1}
\end{table}

\subsection{Role of the hadron wave functions}
\label{sec:hamiltonian2}

In the theoretical description of the production of heavy hadrons in heavy-ion collisions, heavy quarks are predominantly hadronized through coalescence at low transverse momenta, the probability of coalescence decreasing at increasing $p_T$ when fragmentation starts to contribute~\cite{{Greco:2003mm},{Greco:2003xt}}.
This is because when a heavy quark with high transverse momentum is far away from light quarks (antiquarks) in momentum space, the coalescence probability is significantly low. In contrast, a heavy quark with low transverse momentum, surrounded by light quarks (antiquarks) in its neighboring phase space, exhibits a considerably higher coalescence probability. 

In the coalescence model, the spatial extension of the produced hadron plays a significant role. A single-parameter Gaussian shape is usually adopted for the Wigner distribution in space and momentum~\cite{Oh:2009zj,Greco:2004}. Within a quark-model perspective, such a Wigner function is obtained by assuming a single Gaussian for the hadron wave function, which provides a simple effective description of basic properties of the produced hadron.

There exist different strategies in the literature to determine the width parameter of the single-Gaussian wave function of the hadrons. Generally speaking, it is usually required that all charm quarks at zero transverse momentum are hadronized entirely by coalescence. References~\cite{{Cho:2019syk},{Greco:2004},{Oh:2009zj},{Cho:2020},{Cao:2020}} assumed the same Gaussian parameter for all different physical states produced. Such a procedure results in a large value for the size of the $D^0$-meson. For example, Ref.~\cite{Cho:2020} used for the $D^0$-meson
a single-Gaussian wave function with parameter $\omega =0.078$~GeV for RHIC and $\omega =0.076$~GeV at LHC. Thus, using $m_{c} = 1.5$~GeV and $m_{q}=0.3$~GeV, the root-mean-square  (r.m.s.) radius of the $D^0$ meson turns out to be: $r_D^0=\sqrt{3/(2\mu\omega)}=1.75$~fm, which is significantly larger than the presumed physical charge radius of the $D^0$ meson, which is of the order of $\sim1$~fm~\cite{Hwang:2001th,Oh:2009zj,Cho:2019syk}.  

Reference~\cite{Greco:2018} used a different prescription to normalize the coalescence probability. The authors of this reference considered coalescence exclusively in the ground state, but rather than tuning the width parameter of the hadron wave function, they adjusted the coalescence probability through an arbitrary normalization in the Wigner distribution to guarantee that all charm quarks hadronized in the limit $p_T \to 0$, in other words, the normalization goes to unity for zero momentum. However, such a procedure leads to an artificial increase in the baryon-to-meson ratio~\cite{{Cao:2020},{Song:2021mvc}}.

\begin{figure}[htb]
\vspace{-0.6cm}
\begin{center}
\resizebox{11cm}{!}{\includegraphics{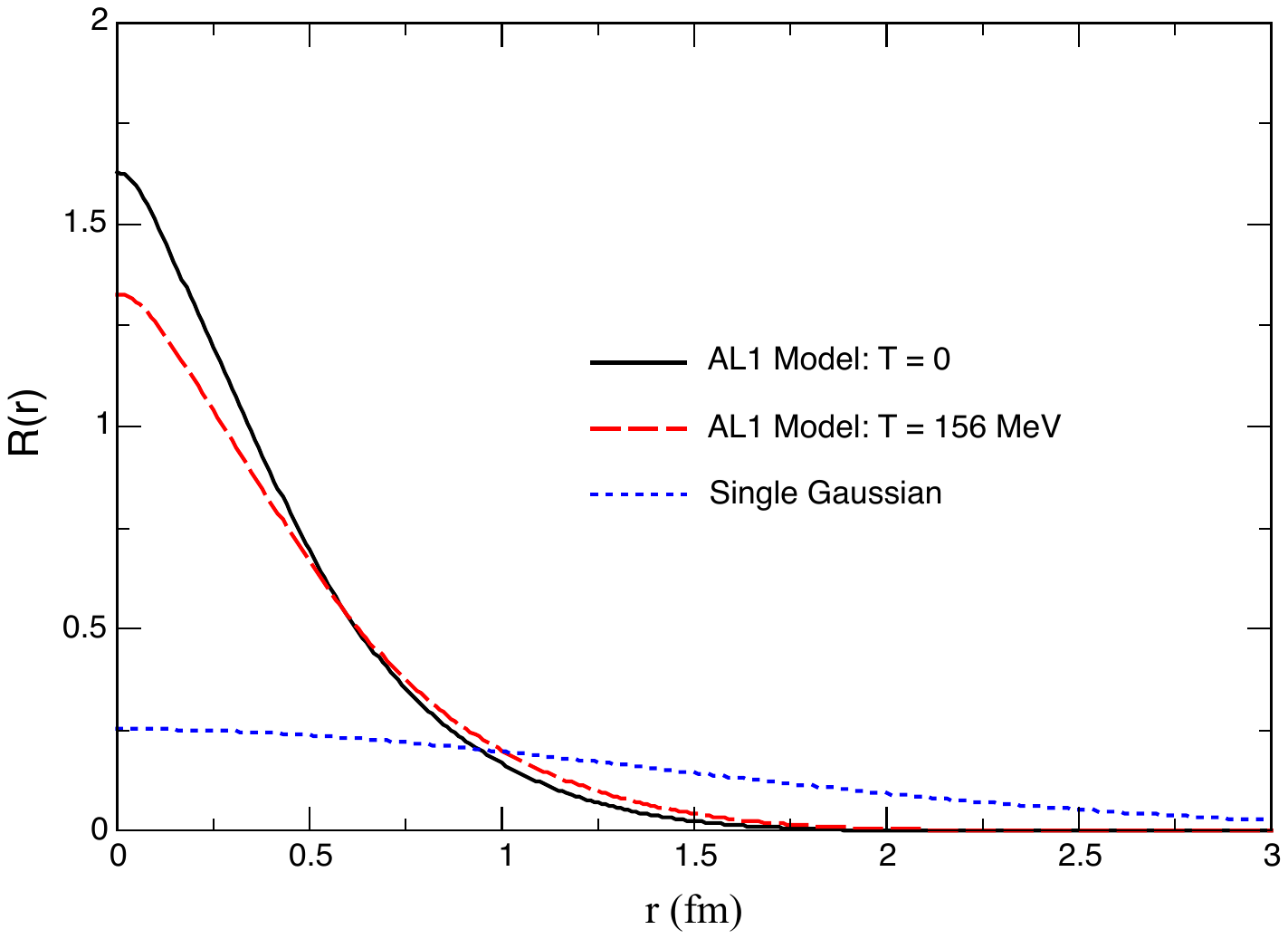}}
\vspace{-0.6cm}
\caption{$D^0$ meson wave function obtained from the AL1 model
at two different temperatures compared to the single-Gaussian form
used in Ref.~\cite{Cho:2020} for the LHC.}
\label{D_wavefunction}
\end{center}
\end{figure}

The stark contrast between single-Gaussian and realistic wave functions is clearly seen in Fig.~\ref{D_wavefunction}, where we compare the $D^0$-meson wave function derived from the AL1 model in both vacuum and at LHC temperatures with the single-Gaussian form used in Ref.~\cite{Cho:2020}. Although one could use a single Gaussian wave function giving a reasonable value for the $D^0$-meson radius, such a wave function cannot capture the more intricate inner structure of the hadron, which is effectively represented by more realistic wave functions. This distinction is equally evident in the case of baryons, as shown in Fig.~2 of Ref.~\cite{Val96}, and is particularly pronounced in the case of tetraquarks, whose wave functions include contributions from different color space vectors compared to those of baryons and mesons, as discussed in Ref.~\cite{Her20}. Consequently, it is imperative to have comparative predictions for the production of various hadron types using realistic wave functions derived from the same model. To address this, the present study aims to minimize uncertainties employing realistic wave functions for all hadrons under investigation, all obtained within the AL1 model. We will revisit normalization issues when comparing the results with the experimental data.

\subsection{Quark transverse momentum distributions}
\label{sec:charm_distribution}

The second important ingredient to evaluate the probability of producing hadrons from quarks in the medium formed in a quark-gluon plasma is the phase-space distribution of the constituents in the medium, that take into account the total number of quarks determined by the initial hard scattering. Thus, the charm and bottom quark distributions are taken from distinct phenomenological models adjusted to reproduce features of particle production in high-energy collisions. For charm quarks, these models are often constrained by experimental data from $pp$ and $AA$ collisions. For bottom quarks, in the absence of direct experimental data for $B$ mesons in Au+Au collisions at RHIC or Pb+Pb collisions at the LHC at $\sqrt{s_{NN}} = 2.76$ TeV, the distributions are derived from theoretical extrapolations or fits to model results, as discussed in Refs.~\cite{Oh:2009zj,Cao:2013ita}. These distributions incorporate medium effects at different temperatures and are calibrated to match overall collision conditions such as energy, volume, and effective temperature.

We use a coalescence approach based on a fireball scenario, where the bulk
of the particles consists of a thermalized system comprising gluons and $u$, $d$, $s$ quarks and their
corresponding antiquarks. It is assumed that the longitudinal momentum distribution remains invariant
under boosts within the rapidity range (in momentum space) $y \in (-0.5, +0.5)$.

For the light-quark momentum distribution, we adopt a fully relativistic thermal distribution,
\be
\frac{d^2N_l}{d^2 p_T}=g_l\frac{V}{(2\pi)^3} \, m_T \, e^{-\frac{m_T}{T_{\rm eff}}} \, ,
\label{dNldpT}
\ee
where $g_l = 6$ is the light-quark color and spin degeneracy, and the transverse mass $m_T$ is given by $m_T=\sqrt{p_T^2+m^2}$ with $m$ being the quark mass. 

Some of the needed transverse momentum distributions of charm and bottom quarks are given by explicit formulas in the literature, and others are available only through figures. In the latter case, we have extracted numerical data from the images and made our own fits. For charm quarks, Plumari {\it et al.}~\cite{Greco:2018} provide a parametrization for Au+Au collisions at RHIC with $\sqrt{s_{NN}} = 200$~GeV, 
\be
\frac{d^2N_c}{d^2 p_{T}}=\left\{
\begin{array}{ll}
0.69\,e^{-1.22\,p_{T}^{1.57}} & \quad p_{T} \le 1.85~\textrm{GeV} \\
1.08\,e^{-3.04\,p_{T}^{0.71}}+3.79\,(\,1.0+p_{T}^{2.02}\,)^{-3.48} & \quad p_{T} > 1.85~\textrm{GeV} \end{array} \right.\, ,
\label{dNcdpT_rhic}
\ee
and for Pb+Pb collisions at LHC with $\sqrt{s_{NN}} = 2.76$~TeV,
\be
\frac{d^2N_c}{d^2 p_{T}}=\left\{
\begin{array}{ll}
1.97\,e^{-0.35p_{T}^{2.47}} & ~p_{T} \le 1.85~\textrm{GeV} \\
7.95\,e^{-3.49\,p_{T}^{3.59}}+87335\,(\,1.0+p_{T}^{0.5}\,)^{-14.31} & ~p_{T} > 1.85~\textrm{GeV}\\
\end{array} \right. \, .
\label{dNcdpT_lhc276}
\ee
For Pb+Pb collisions at LHC with $\sqrt{s_{NN}} = 5.02$~TeV, we extracted the
data from Fig.~1 of Minissale {\it et al.}~\cite{Minissale:2023dct} and fitted them with the following expression:
\be
\frac{d^2N_c}{d^2 p_{T}}=\,\frac{2.42}{\left(1 + 0.19 \, p_{T}^{2.05}\right)^{3.24}} .
\label{dNcdpT_lhc502}
\ee

For bottom quarks in Au+Au collisions at RHIC with 
$\sqrt{s_{NN}} = 200$ GeV, we use the parametrization given by Oh {\it et al.} in eq.~(28) of Ref.~\cite{Oh:2009zj}: 
\be
\frac{d^2N_b}{d^2 p_{T}} = 0.0025\Big{[}\,1\,+\,\left(\frac{p_{T}}{16}\,\right)^{5}\,\Big{]}\,e^{- 0. 67 p_{T}} \, .
\label{dNbdpT_rhic}
\ee
For Pb+Pb collisions at the LHC with $\sqrt{s_{NN}} = 2.76$~TeV, we fit the results shown in Fig.~5(b) 
of Ref.~\cite{Cao:2013ita}:
\be
\frac{d^2N_b}{d^2 p_{T}}=\frac{0.013 \,p^{2}_{T}\,\,e^{-0.64\,p_{T}}}{\left(1+0.32\,\,p_{T}\right)^{2}} .
\label{dNbdpT_lhc276}
\ee
Finally, for Pb+Pb collisions at the LHC with $\sqrt{s_{NN}} = 5.02$~TeV, we fit the results shown in Fig.~16 (top left panel) of Ref.~\cite{Zhao:2024ecc}:
\be
\frac{d^2N_b}{d^2 p_{T}} = \frac{0.48\,p_{T}^{1.03}}{129.07 + p_{T}^{3.84}} \, .
\label{dNbdpT_lhc502}
\ee

We take volumes $V = 2100$~fm$^3$ for RHIC~\cite{And13} and $V = 5380$~fm$^3$ for LHC~\cite{Sta14}, and the same effective temperature $T_{\rm eff} = 156$~MeV~\cite{And13,Sta14} for both RHIC and LHC. The choice of parameters assumes
that the critical temperature representing the transition from the initial QGP to the final hadronic matter is identical to the temperature of hadronization. Moreover, it assumes the equivalence of the critical volume and the volume of hadronization. During the crossover phase transition, there is continuous production of hadrons that requires the equality between the number of hadrons generated by statistical hadronization and the number produced through the coalescence of the constituent quarks at the endpoint of the phase~\cite{Cho:2017dcy}.

The hadronization temperature is a parameter that is
obtained by fitting experimental data by means of a model. The statistical
hadronization model is commonly used to derive, among others, the
hadronization temperature. The best fit of the CERN LHC data in 
Ref.~\cite{Sta14}
yields a temperature of 156 MeV, slightly below the expectation
from RHIC data of about 164 MeV~\cite{And13}. 
These determinations are not free
of uncertainties and the value of 164~MeV is obtained by removing protons
and antiprotons from the fit; in case they are included, a value of 
152~MeV is obtained. What is clearly specified is that the extracted
temperature values, which first increase sharply with increasing beam 
energy, level off near 160~MeV for energies $\sqrt{s_{NN}}> 20$~ GeV.
This is the reason why we choose the value that we considered best established, $T=156$~MeV.
A common value ensures a uniform and consistent framework for analyzing the results of the different collisions at RHIC and LHC.
That being said, such small temperature variations in 
the coalescence model generate very small changes in the 
results, less than 5\%, since the parameters affecting the 
heavy quarks do not vary with temperature and those of the light 
quarks do so very smoothly for small temperature variations.
However, it should be kept in mind that although 
small variations in the hadronization temperature may lead to negligible 
changes in the spectra, they can significantly affect the elliptic 
flow~\cite{Rap:2018,Rap:2024}. Therefore, it would be desirable to have heavy 
quark distributions evaluated with a single model and the same
hadronization temperature.

As discussed above, effective light quark masses are influenced by chiral symmetry restoration ($\chi\rm{SR})$, whereas those of heavy quarks are not. Therefore, $\chi\rm{SR}$ will affect the transverse momentum distributions, since they depend on the quark masses. For the light quarks we consider two scenarios, one that neglects $\chi\rm{SR}$ and the other that takes it into account. 
In the first scenario, $m_{u,d} = 0.315$~GeV and in the second $m_{u,d} = 0.221$ GeV, as shown in Table~\ref{Tab1}. Therefore, integrating Eq.~\eqref{dNldpT}, we obtain the following numbers of light quarks (first scenario results indicated by No $\chi{\rm SR}$ and second scenario ones by $\chi{\rm SR}$):
\begin{equation}
\begin{array}{lccccccc}
&  & {\rm RHIC}   &     &\hspace{1.5cm}       &          & {\rm LHC} &  \\[-0.1true cm]
   & {\rm No}\; {\chi{\rm SR}} & &  \chi{\rm SR} &  & {\rm No}\; {\chi{\rm SR}} &  & \chi{\rm SR}\\
   \hline\\[-0.6true cm]
N_{u,d} \simeq & 188 &  & 262  & & 542 & & 670
\end{array}
\label{lq-numbers}
\end{equation}
For the heavy quarks, using vacuum quark mass values, we obtain integrating Eqs.~\eqref{dNcdpT_rhic}--\eqref{dNbdpT_lhc502}, the following total numbers of charm and bottom quarks:

\begin{equation}
\begin{array}{lccc}
& \ {\rm RHIC}\,(0.2\,{\rm TeV}) \ & {\rm LHC}\,(2.76\,{\rm TeV}) \ & \ {\rm LHC}\,(5.02\,{\rm TeV}) \\
\hline \\[-0.6true cm]
N_c \simeq  & 2    & 14   & 15 \\
N_b \simeq & 0.034 & 0.44 & 0.74
\end{array}
\label{hq-numbers}
\end{equation}

\subsection{Heavy hadron transverse momentum distributions}
\label{sec:hadron_distribution}

In the derivation of Eq.~\eqref{3CoalTrans} it is assumed a uniform spatial distribution for quarks within the hadronized QGP characterized by cylindrical volume~$V$. Using the Gaussian form of the Wigner function of Sect.~\ref{sec:Wigner}, one can perform the integrations over the spatial components of the coalescence formula that leads to the transverse momentum spectrum of a given type of hadrons.  For mesons, the expression we need to integrate is the following:
\be
\frac{d^2N_{M}}{d^2\bp_T}=\frac{g_{M}}{V}
\int\,d^2\bp_{QT}\,d^2\bp_{\bar{l}T}\,\delta^{(2)}\left(\bp_{T}-\bp_{QT} - \bp_{\bar{l}T}\right)\, \frac{d^2N_{Q}}{d^2\bp_{QT}}\,\frac{d^2N_{\bar{l}}}{d^2\bp_{\bar{l}T}}\,\rho_{M}\left(\bk\right)\,, \label{D0CoalTrans}
\ee
where $g_M$ is the degeneracy factor for a colored quark and antiquark to form a color-neutral meson $D^{0}\,(\bar{u}c)$ and $B^{0}\,(\bar{u}b)$, i.e., $g_{D^{0}}=g_{B^{0}}=1/(3^2\times2^2)$. $\rho_{M}$ is the Wigner function given in Eq.~\eqref{wigner-jacobi-meson}, and $\bk$ is the relative quark transverse momentum defined in Eq.~\eqref{meson-rel-mom} in terms of the momentum of the heavy quark (\,$\bp_{Q}$\,) and the light antiquark (\,$\bp_{\bar{l}}$\,). 

The transverse momentum distribution of a heavy quark and two light quarks to form $\Lambda_{Q_{1}}\,(udQ_{1})$, $\Sigma_{Q_{1}}\,(udQ_{1})$, and $\Xi_{Q_{1}}\,(usQ_{1})$ baryons is given by:
\bea
\frac{d^2N_{B_{Q_{1}}}}{d^2\bp_T}&=&
\frac{g_{B_{Q_{1}}}}{V^{2}}\,\int\,d^2\bp_{Q_{1}T}\,d^2\bp_{l_{1}T}\,d^2\bp_{l_{2}T}\,\delta^{(2)}\left(\bp_{T}-\bp_{Q_{1}T} - \bp_{{l_{1}}T}- \bp_{{l_{2}}T}\right)
\nonumber\\[0.1cm]&\times&
\frac{d^2N_{Q_{1}}}{d^2\bp_{Q_{1}T}}\, \frac{d^2N_{l_{1}}}{d^2\bp_{l_{1}T}}\,
\frac{d^2N_{l_{2}}}{d^2\bp_{l_{2}T}}\,\rho_{B_{Q_{1}}}\left(\bk_{1},\bk_{2}\right) \, ,
\label{qqQCoalTrans}
\eea
where the index $Q_{1}$ refers to the heavy quark ($c$ or $b$) and indices $l_{1}$ and $l_2$ refer to the light quarks ($u$, $d$ or $s$). $g_{B_{Q_{1}}}$ is the statistical factor: $g_{\Lambda_{Q_{1}}}=2/(3^3\times2^3)$, $g_{\Sigma_{Q_{1}}}=(2\times3)/(3^3\times2^3)$, and $g_{\Xi_{Q_{1}}}=(2\times2)/(3^3\times2^3)$. The relative transverse momenta $\bk_1$ and $\bk_2$ are defined in Eq.~\eqref{bar-rel-mom1}, and the Wigner function $\rho_{B}$ is given in Eq.~\eqref{wigner-jacobi-baryon}. The transverse momentum distribution of two heavy quarks and a light quark to form a $\Xi_{Q_1Q_2}\,(uQ_1Q_2)$ baryon is given by:
\bea
\frac{d^{2}N_{_{B_{Q_1Q_2}}}}{d^2\bp_T}&=&
\frac{g_{_{B_{Q_1Q_2}}}}{V^{2}}\,\int\, d^2\bp_{Q_{1}T}\, d^2\bp_{Q_{2}T}\, d^2\bp_{l_{1}T}\,\delta^{(2)}\left(\bp_{T}-\bp_{Q_{1}T} - \bp_{{Q_{2}}T}- \bp_{{l_{1}}T}\right)
\nonumber\\[0.1cm] &\times &\frac{d^2N_{Q_{1}}}{d^2\bp_{Q_{1}T}}\,\frac{d^2N_{Q_{2}}}{d^2\bp_{Q_{2}T}} \, \frac{d^2N_{l_{1}}}{d^2\bp_{l_{1}T}}
\,\rho_{_{B_{Q_1Q_2}}}\left(\bk_{1},\bk_{2}\right),
\label{qQQCoalTrans}
\eea
where the indices $Q_{1}$ and $Q_{2}$ refer to the heavy quarks and the index $l_{1}$ refers to light quarks. $g_{B_{Q}}$ is the statistical factor, $g_{\Xi_{Q_1Q_2}}=(2\times2)/(3^3\times2^3)$. The relative transverse momenta $\bk_1$ and $\bk_2$ are defined in Eq.~\eqref{bar-rel-mom1}, and the Wigner function  $\rho_{B}$ is given in Eq.~\eqref{wigner-jacobi-baryon}.

Finally, the transverse momentum distribution of a $T_{Q_1Q_2}\,(Q_1Q_2\bar{u}\bar{d})$ tetraquark obtained from two heavy quarks, $Q_i = c$ \text{or} $b$, and two light quarks, $l_i=\bar{u}$ and $\bar{d}$ is given by:
\bea
\frac{d^{2}N_{_{T_{Q_1Q_2}}}}{d^2\bp_T}&=&
\frac{g_{_{T_{Q_1Q_2}}}}{V^{3}}\,\int\,d^2\bp_{Q_{1}T}\,d^2\bp_{Q_{2}T}\,d^2\bp_{\bar{l}_{1}T}\,\,d^2\bp_{\bar{l}_{2}T}\,\delta^{(2)}\left(\bp_{T}-\bp_{Q_{1}T} - \bp_{{Q_{2}}T}- \bp_{{\bar{l}_{1}}T}-\bp_{{\bar{l}_{2}}T}\right)
\nonumber\\[0.1cm] &\times &
\frac{d^2N_{Q_{1}}}{d^2\bp_{Q_{1}T}}\, \frac{d^2N_{Q_{2}}}{d^2\bp_{Q_{2}T}}\, \frac{d^2N_{\bar{l}_{1}}}{d^2\bp_{\bar{l}_{1}T}}\, \frac{d^2N_{\bar{l}_{2}}}{d^2\bp_{\bar{l}_{2}T}}\,\rho_{_{T_{Q_1Q_2}}}\left(\bk_{1},\bk_{2},\bk_{3}\right),
\label{TQQCoalTrans}
\eea
where $g_{T_{Q_1Q_2}}= 3/(2^4 \times 3^4)$ is the statistical factor giving the probability of the formation of a $T_{Q_1Q_2}$ tetraquark from two heavy quarks, $Q_{1}$ and $Q_{2}$, and two light quarks, $l_{1}$ and $l_{2}$. The relative transverse momenta $\bk_1$, $\bk_2$ and $\bk_3$ are defined in Eq.~\eqref{tetra-rel-mom1} and the Wigner function $\rho_T$  
is defined in Eq.~\eqref{wigner_final_tetratquark}.
 
\section{Results}
\label{sec:5}
\subsection{Yields}
\label{yields}
We present results for the yields of open-charm and bottom hadrons, as well as exotic tetraquarks: $D^0$, $\Lambda_{c}$, $\Sigma_{c}$, $\Xi_{c}$, $\Xi_{cc}$, $T_{cc}$, $B^{0}$, $\Lambda_{b}$, $\Sigma_{b}$, $\Xi_{b}$, $\Xi_{bc}$, $\Xi_{bb}$, $T_{bc}$, and $T_{bb}$.
We also investigate the effect of chiral symmetry restoration on this observable, through a finite temperature. As discussed above, finite temperature due to the dynamical breaking of chiral symmetry 
influences the parameters of the interacting potential, specifically the mass of the constituent light quark. The temperature dependence of those parameters have been obtained using predictions of the Nambu--Jona-Lasinio model~\cite{{Nam61}}, following the strategy set up in our work in Ref.~\cite{Carames:2016qhr}, in which the effects of a hot and dense medium on the binding energy of hadronic molecules with open-charm mesons were studied.

The yields are obtained by integrating Eqs.~\eqref{D0CoalTrans}--\eqref{TQQCoalTrans} and using the hadronization temperature and volume discussed in the previous section. Table~\ref{Tab2} presents our results for the yields for central Au+Au and Pb+Pb collisions at RHIC and LHC energies. The results labeled ${\rm No}\;{\chi{\rm SR}}$ correspond to the scenario in which chiral symmetry restoration effects on the coalescence process are not considered and those labeled ${\chi{\rm SR}}$ take into account chiral restoration effects.

\begin{table}[h]
\caption{Open-flavor heavy meson, baryon and tetraquark yields at mid-rapidity in the coalescence model
expected at RHIC at $\sqrt{s_{NN}}=0.2$ TeV Au+Au collisions, at LHC at $\sqrt{s_{NN}}=2.76$~TeV and $\sqrt{s_{NN}}=5.02$~TeV Pb+Pb collisions.   All results are calculated using the AL1 model hadron wave functions.}
\begin{ruledtabular}
\begin{tabular}{ccccccc}
&\multicolumn{2}{c}{RHIC ($0.2$~TeV)} &\multicolumn{2}{c}{LHC ($2.76$~TeV)}
&\multicolumn{2}{c}{LHC ($5.02$~TeV)}  \\
\cline{2-3} \cline{4-5} \cline{6-7}
&No\;${\chi{\rm SR}}$& $\chi{\rm SR}$ &No\;${\chi{\rm SR}}$ & $\chi{\rm SR}$
&No\;${\chi{\rm SR}}$ & $\chi{\rm SR}$\\
\hline
\hline
$N_{{D}^{0}}$&$4.2\times10^{-2}$& $6.6\times10^{-2}$&$2.9\times10^{-1}$&$4.7\times10^{-1}$&$3.1\times10^{-1}$& $5.1\times10^{-1}$
\\
$N_{B^{0}}$     &$6.6\times10^{-4}$ & $1.1\times10^{-3}$  & $7.3\times10^{-3}$  & $1.3\times10^{-2}$  & $1.4\times10^{-2}$  & $2.4\times10^{-2}$ 
\\
\hline
$N_{\Lambda_{c}}$&$3.1\times10^{-2}$&$6.5\times10^{-2}$ &$2.0\times10^{-1}$ &$4.4\times10^{-1}$&$2.1\times10^{-2}$&$4.7\times10^{-1}$
\\
$N_{\Sigma_{c}}$&$1.2\times10^{-1}$&$2.9\times10^{-1}$&$7.7\times10^{-1}$&$1.9$&$8.2\times10^{-1}$&$2.1$
\\
$N_{\Xi_{c}}$&$1.8\times10^{-2}$ &$2.6\times10^{-2}$&$1.1\times10^{-1}$&$1.7\times10^{-1}$&$1.1\times10^{-1}$&$1.7\times10^{-1}$
\\
$N_{\Xi_{cc}}$&$4.7\times10^{-5}$&$7.5\times10^{-5}$&$5.8\times10^{-4}$&$9.4\times10^{-4}$&$6.2\times10^{-4}$&$9.9\times10^{-4}$ 
\\
\hline
$N_{\Lambda_{b}}$ &$3.6\times10^{-4}$ & $8.2\times10^{-4}$  & $3.6\times10^{-3}$   & $8.6\times10^{-3}$  & $6.3\times10^{-3}$  & $1.6\times10^{-2}$ 
\\
$N_{\Sigma_{b}}$  &$1.5\times10^{-3}$ & $3.8\times10^{-3}$  & $1.5\times10^{-2}$  & $4.0\times10^{-2}$  & $2.5\times10^{-2}$  & $7.4\times10^{-2}$ 
\\
$N_{\Xi_{b}}$     &$1.8\times10^{-4}$ &$2.7\times10^{-4}$   & $1.6\times10^{-3}$  & $2.5\times10^{-3}$  & $2.7\times10^{-3}$  & $4.3\times10^{-3}$ 
\\
$N_{\Xi_{bc}}$   &$1.7\times10^{-7}$ & $2.4\times10^{-7}$  & $3.8\times10^{-7}$  & $5.2\times10^{-7}$  & $4.3\times10^{-7}$  & $6.0\times10^{-7}$ 
\\
$N_{\Xi_{bb}}$   &$1.3\times10^{-9}$ & $2.0\times10^{-9}$  & $2.8\times10^{-8}$  & $3.9\times10^{-8}$  & $7.2\times10^{-8}$  & $9.2\times10^{-8}$ 
\\
\hline
$N_{T_{cc}}$&$1.1\times10^{-5}$&$2.7\times10^{-5}$&$8.1\times10^{-5}$&$1.9\times10^{-4}$&$9.4\times10^{-5}$&$2.2\times10^{-4}$ 
\\
$N_{T_{bc}}$   &$1.9\times10^{-8}$& $3.2\times10^{-8}$  & $1.1\times10^{-7}$  & $2.2\times10^{-7}$  & $1.5\times10^{-7}$  & $3.0\times10^{-7}$
\\
$N_{T_{bb}}$   &$9.8\times10^{-10}$& $1.2\times10^{-9}$  & $6.7\times10^{-9}$  & $8.9\times10^{-9}$  & $8.3\times10^{-9}$  & $1.1\times10^{-8}$
\\
\end{tabular}
\end{ruledtabular}
\label{Tab2}
\end{table}

The first important result that we highlight is that in a realistic model the yields of exotic hadrons, in particular the $T_{QQ}$ tetraquarks, are of the same order of magnitude as the corresponding doubly-heavy baryons. This can be understood in terms of the internal structure of the $QQ\bar u\bar d$ tetraquarks. In the limit $M_Q/m_q \gg 1$ the heavy quark pair is dominantly in a spin~1 state, as it occurs in a $\Xi_{QQ}$ baryon, the main difference in the yields arising from the degeneracy factors. For the $T_{QQ^\prime}$ tetraquarks, the comparison between the $J^P=1^+$ tetraquark and the $\Xi_{bc}(1/2^+)$ baryon is also almost direct. Thus,  when the strategy for the detection of doubly heavy baryons is improved in current accelerators~\cite{Aaij:2017}, an equivalent method for the detection of tetraquarks will be available through an increase in energy.

Regarding temperature effects, our results show that partial chiral symmetry restoration enhances the yields, being more pronounced for the lightest hadrons $D^0$, $B^0$, $\Lambda_{c}$, $\Lambda_{b}$, $\Sigma_{c}$, and $\Sigma_{b}$. This enhancement is due to two different factors. Firstly, the root mean square (r.m.s.) radii of these particles increase, as illustrated in Fig.~\ref{D_wavefunction} for the $D^0$ meson, thereby enhancing the coalescence probability. Secondly, the number of light quarks also increases with temperature, contributing to the enhancement of the yields.
However, the effects of partial chiral symmetry restoration on the spatial distribution of heavier hadrons, such as $\Xi_{c}$, $\Xi_{cc}$, $\Xi_{b}$, $\Xi_{bc}$, $\Xi_{bb}$, are less pronounced.
The impact of partial chiral symmetry on the yields of $T_{cc}$, 
$T_{bc}$, and $T_{bb}$ is small. Although temperature significantly affects their spatial distribution, resulting in a decrease of the r.m.s. radius
of the $T_{cc}$ from $0.51$~fm at $T=0$ to $0.36$~fm at $T=156$~MeV, this effect is practically offset by the increase in the number of light quarks, resulting in negligible changes in the yields.

Another result we highlight is that the yields obtained using a single Gaussian wave function with a fitted width parameter~\cite{Cho:2019syk, Cho:2020, Greco:2018} are nearly one order of magnitude larger than those obtained with realistic wave functions. The reason for this difference is rather simple. Recall that either the width of the Gaussian wave function used in these calculations leads to hadrons sizes larger than expected, or a normalization factor must be included in the Wigner function to mimic the experimental data. Both assumptions make it possible that all heavy quarks at zero transverse momentum get hadronized by quark coalescence.  Hadrons are produced via two different mechanisms in heavy-ion collisions,  by quark coalescence and fragmentation~\cite{Fries:2008hs,Greco:2003xt,Greco:2003mm}. Since the hadron production by quark coalescence is dominant at low transverse momenta, it is natural to expect exclusive hadron production at zero transverse momentum by quark coalescence, enhancing the predicted yields. In our results, it is important to emphasize that the increase in yields is due to the use of a fully relativistic expression for the light-quark distribution, Eq.~\eqref{dNldpT}. If we compare with the results of Ref.~\cite{Fontoura:2019opw}, in which we used a non-relativistic approximation for the light quark distribution, we observe an increase by roughly a factor~2, mainly due to the increase of the number of light quarks.

Finally, regarding the baryon-to-meson ratio, we obtain $N_{\Lambda_c}/N_{D^0}=0.74$ at RHIC without chiral symmetry restoration effects and $0.98$ considering chiral symmetry restoration effects. Our results considering chiral symmetry restoration effects, which we believe are the most realistic ones, are consistent with the recent results by the STAR Collaboration, $N_{\Lambda_c}/N_{D^0}=1.08 \pm 0.16 ({\rm stat.}) \pm 0.26 ({\rm sys.})$~\cite{STAR2020}. It is important to note that this result is significantly larger than the PYTHIA model calculations for p+p collisions. For LHC~\cite{Acharya:2018,Acharya:2019} at 2.76~TeV, we obtain $N_{\Lambda_c}/N_{D^0}=0.69$ without chiral symmetry restoration effects and $0.94$ considering chiral symmetry restoration effects, and at 5.02~TeV, we get $N_{\Lambda_c}/N_{D^0}=0.68$ without chiral symmetry restoration effects and $0.92$ considering chiral symmetry restoration effects. For the bottom sector, considering chiral symmetry restoration effects we get
$N_{\Lambda_b}/N_{B^0}=0.75$ at RHIC,
$N_{\Lambda_b}/N_{B^0}=0.66$ at LHC at 2.76~GeV, and
$N_{\Lambda_b}/N_{B^0}=0.45$ at LHC at 5.02~GeV.

\subsection{Production spectra for hadrons with charm and bottom}
\label{EduardoPL_sec:4}
\begin{figure}[h!]
\begin{center}
\makebox[\textwidth][l]{\hspace{-0.7cm}\includegraphics[trim=1 1 1 90, clip,width=1.26\textwidth]{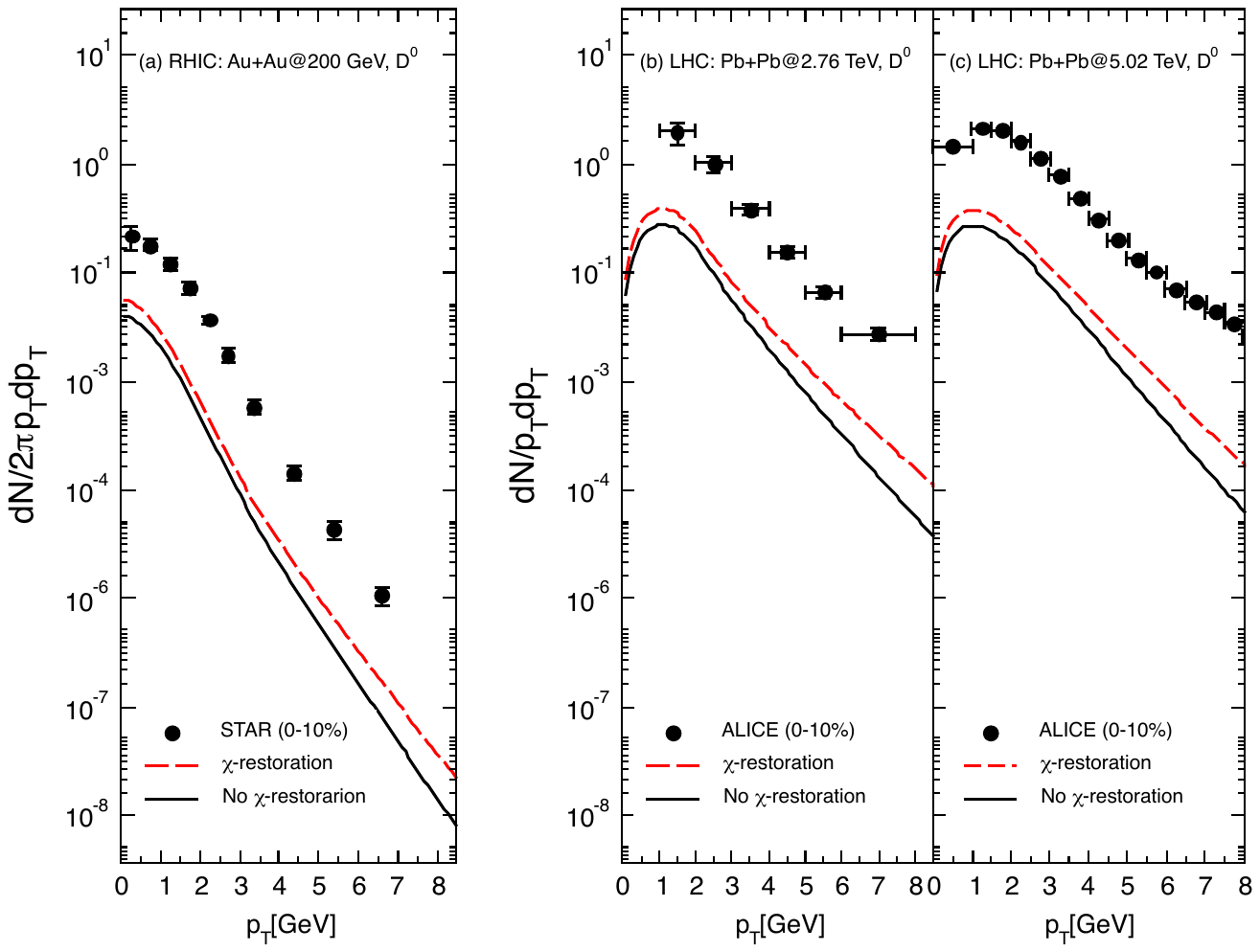}}
\\[-1.0true cm]
\caption{Transverse momentum spectra for $D^0$ meson production using the AL1 model wave functions at different collision energies: RHIC ($200$~GeV) and LHC ($2.76$~TeV and $5.02$~TeV).
No $\chi$-restoration results are denoted by the black solid line and $\chi$-restoration results by the red dashed line. Experimental data are taken from Refs.~\cite{{STAR:2018zdy},{Alice:2012zdy},{Raa2020}}.}
\label{spect_d0}
\end{center}
\end{figure}

In Fig.~\ref{spect_d0}, we present the transverse momentum spectra for $D^0$ production at RHIC ($200$~GeV) and LHC ($2.76$~TeV, $5.02$~TeV). The spectra are calculated using the realistic AL1 model hadron wave functions in the two scenarios discussed above: without (black solid line) and with (red dashed line) $\chi$-restoration effects at the coalescence temperature. These results are compared to experimental data from the STAR and ALICE Collaborations~\cite{{STAR:2018zdy},{Alice:2012zdy},{Raa2020}} for the corresponding collision energies, shown as black circles with error bars.

The experimental data at both temperatures at the LHC~\cite{Alice:2012zdy,{Raa2020}} show smooth spectra, with a peak around $p_T \approx 1-2$~GeV.
The black solid lines (no $\chi$-restoration effects)  underestimate the data at intermediate $p_T$. However, the red dashed lines, considering $\chi$-restoration effects, improve the agreement and closely match the
experimental data up to $p_T=6$~GeV,
except for a normalization factor, as discussed above. The deviation between theoretical calculations and experimental data becomes more apparent at higher $p_T$, indicating the increasing importance of mechanisms beyond chiral symmetry breaking effects, as it may be fragmentation~\cite{Fries:2008hs,Greco:2003xt,Greco:2003mm}. 

\begin{figure}[t]
\begin{center}
\makebox[\textwidth][c]{\includegraphics[trim=1 1 1 90, clip,width=1.2\textwidth]{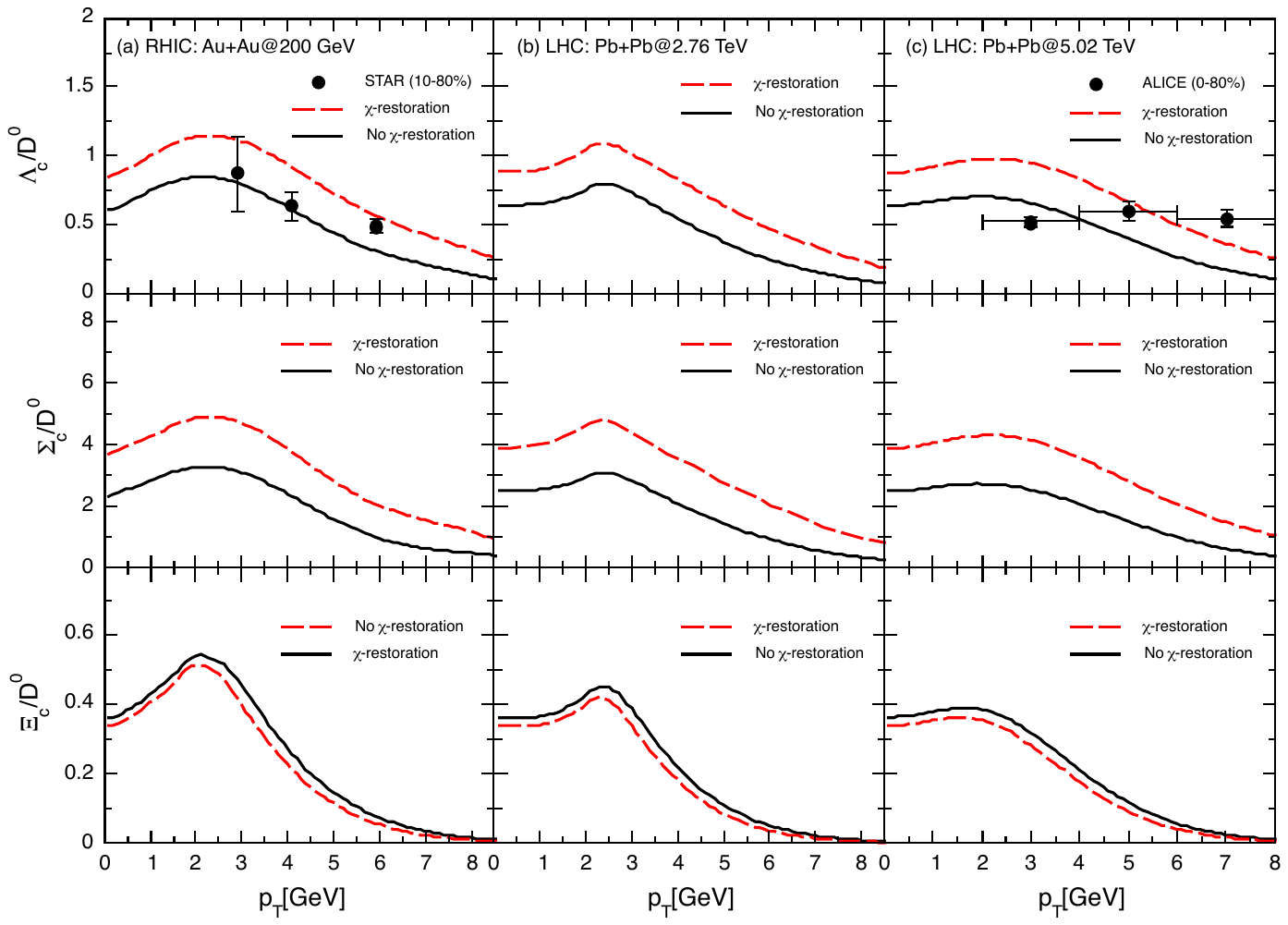}}
\\[-1.0true cm]
\caption{Transverse momentum distribution ratios between $\Lambda_{c}/D^{0}$, $\Sigma_c/D^0$ and $\Xi_c/D^0$ at RHIC and LHC using the AL1 model wave functions. No $\chi$-restoration are denoted by the black solid line and $\chi$-restoration results by the red dashed line. Experimental data are taken from Ref.~\cite{Raa2020}.}
\label{ratio_charm}
\end{center}
\end{figure}

The spectra are found to be one order of magnitude lower than the experimental measurements at low transverse momenta. As discussed above, a global normalization is included either by fitting the width of the single-Gaussian wave functions, or by using a global normalization factor in the Wigner distribution~\cite{Cho:2019syk,Oh:2009zj,Greco:2018}.
We preferred keep the absolute value and look to the distribution
functions that closely agree with the shape of the experimental data up to some intermediate value of the transverse momentum. As also mentioned above, not all heavy quarks in the quark-gluon plasma participate in coalescence; at high energies, some undergo fragmentation, especially relevant for heavy flavors like charm and bottom quarks. There is a lack of consensus on the transverse momentum values at which fragmentation becomes significant for the final production spectra, which may explain the tiny deviation from the experimental trend at high transverse momenta. 
Our main goal is to present a unified picture of the production of mesons,
baryons, and tetraquarks in a coalescence model based on realistic wave functions, including temperature effects, in an accurate manner. Thus,  to avoid ambiguities associated with the absolute value of the transverse momentum spectra, from now on we present the production hadron-to-meson ratio spectra. At this point it is important to emphasize that the results on the first row of 
Fig.~\ref{ratio_charm} clearly show that any normalization 
scheme used to fit
the absolute value of the $D^0$ transverse momentum spectra
would lead to a correct description of the absolute value
of the $\Lambda_c$ transverse momentum spectra.
This is also supported by the baryon-to-meson 
$N_{\Lambda_c}/N_{D^0}$ ratio we have obtained
in Sect.~\ref{yields}, in close agreement
to the results by the STAR Collaboration~\cite{STAR2020}.

Figure~\ref{ratio_charm} shows the transverse momentum spectra of the baryon-to-meson ratios $\Lambda_c/D^0$, $\Sigma_c/D^0$, and $\Xi_c/D^0$. In RHIC Au+Au collisions at $\sqrt{s_{NN}} = 200$~GeV, the $\Lambda_c/D^0$ ratio peaks at $p_T \sim 2.5$~GeV, reaching 1.14 with partial chiral symmetry restoration and 0.84 without it. This result is in line with the recent findings of the STAR Collaboration, which reported a meson-to-baryon ratio of $1.3 \pm 0.5$ for $3 < p_T < 6$~GeV. Furthermore, chiral symmetry restoration effects significantly enhance the $\Sigma_c/D^0$ ratio, particularly around $p_T \approx 3$ GeV, while the $\Xi_c/D^0$ ratio shows an opposite trend, with chiral symmetry restoration effects favoring meson production.

In Pb+Pb collisions at the LHC at $2.76$~TeV, the trends observed at RHIC remain, although with higher baryon-to-meson ratios. Chiral restoration affects the $\Lambda_c/D^0$ and $\Sigma_c/D^0$ ratios over a broader $p_T$ range, while its effect on the $\Xi_c/D^0$ ratio is less pronounced and opposite to that of the $\Lambda_c/D^0$ and $\Sigma_c/D^0$ ratios. At 5.02 TeV, these patterns continue, with the $\Lambda_c/D^0$ and $\Sigma_c/D^0$ ratios showing an increasing trend with increasing temperature. This can be attributed to chiral symmetry restoration, which significantly reduces light-quark masses from $m_{u,d} = 315$~ MeV in vacuum to $m_{u,d} = 221$~ MeV at $T = 156$~MeV. Consequently, $\Lambda_c$ and $\Sigma_c$ baryons, which contain these light quarks, become lighter and more abundantly produced relative to the $D^0$ mesons, whose mass slightly increases with temperature~\cite{Carames:2016qhr}.
In contrast, the $\Xi_c/D^0$ production ratio remains relatively stable with increasing temperature, as the $\Xi_c$ baryon contains a strange quark, less sensitive to temperature-induced mass reductions. The presence of a strange quark mitigates the effects of chiral symmetry restoration on $\Xi_c$ baryons, resulting in negligible changes in their production ratios compared to baryons containing only light quarks.

\begin{figure}[h!]
\begin{center}
\makebox[\textwidth][c]{\includegraphics[trim=1 1 1 90, clip,width=1.2\textwidth]{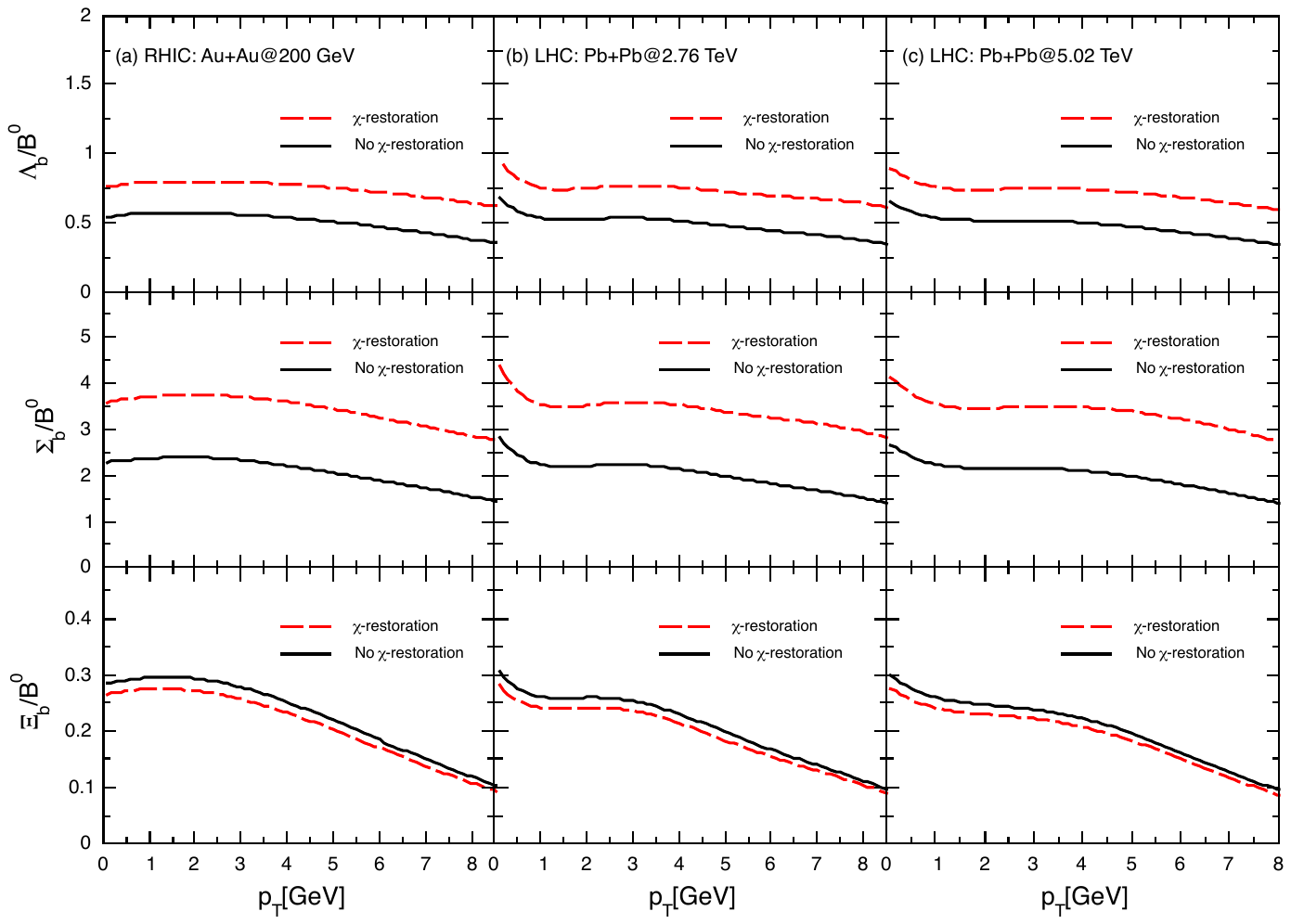}}
\\[-1.0true cm]
\caption{Same as Fig.~\ref{ratio_charm} for $\Lambda_{b}/B^{0}$, $\Sigma_b/B^0$ and $\Xi_b/B^0$.}
\label{ratio_bottom}
\end{center}
\end{figure}

In Figure~\ref{ratio_bottom} we present the results for the transverse momentum spectra of the baryon-to-meson ratios $\Lambda_b/B^0$, $\Sigma_b/B^0$, and $\Xi_b/B^0$.

In both cases, for RHIC and LHC energies (200~GeV, 2.76~TeV, and 5.02~TeV), the spectra show a systematic suppression in the baryon-to-meson ratio as transverse momentum increases. At low transverse momenta, the chiral restoration scenario generally predicts higher ratios than the case without chiral restoration, aligning with theoretical expectations that chiral symmetry restoration enhances baryon production.

A comparison of Figures~\ref{ratio_charm} and~\ref{ratio_bottom}
shows how the ratios for systems containing bottom quarks are lower
than those for systems with charm quarks. This simply reflects the fact that the production of baryons containing bottom quarks is less favorable compared to those with charm quarks because of the larger mass of the bottom quark.
The difference between the chiral restoration and no chiral restoration scenarios is less pronounced in systems with bottom quarks. While chiral restoration significantly increases the ratios in charm systems, the differences are more modest in the bottom-quark sector. The $\Xi_b/B^0$ ratio shows the same stable temperature pattern which appeared in the charm sector. These findings underscore the intricate interplay between quark mass dynamics and hadron production mechanisms in high-temperature QCD environments.

\begin{figure}[t]
\begin{center}
\makebox[\textwidth][c]{\includegraphics[trim=1 1 1 90, clip,width=1.1\textwidth]{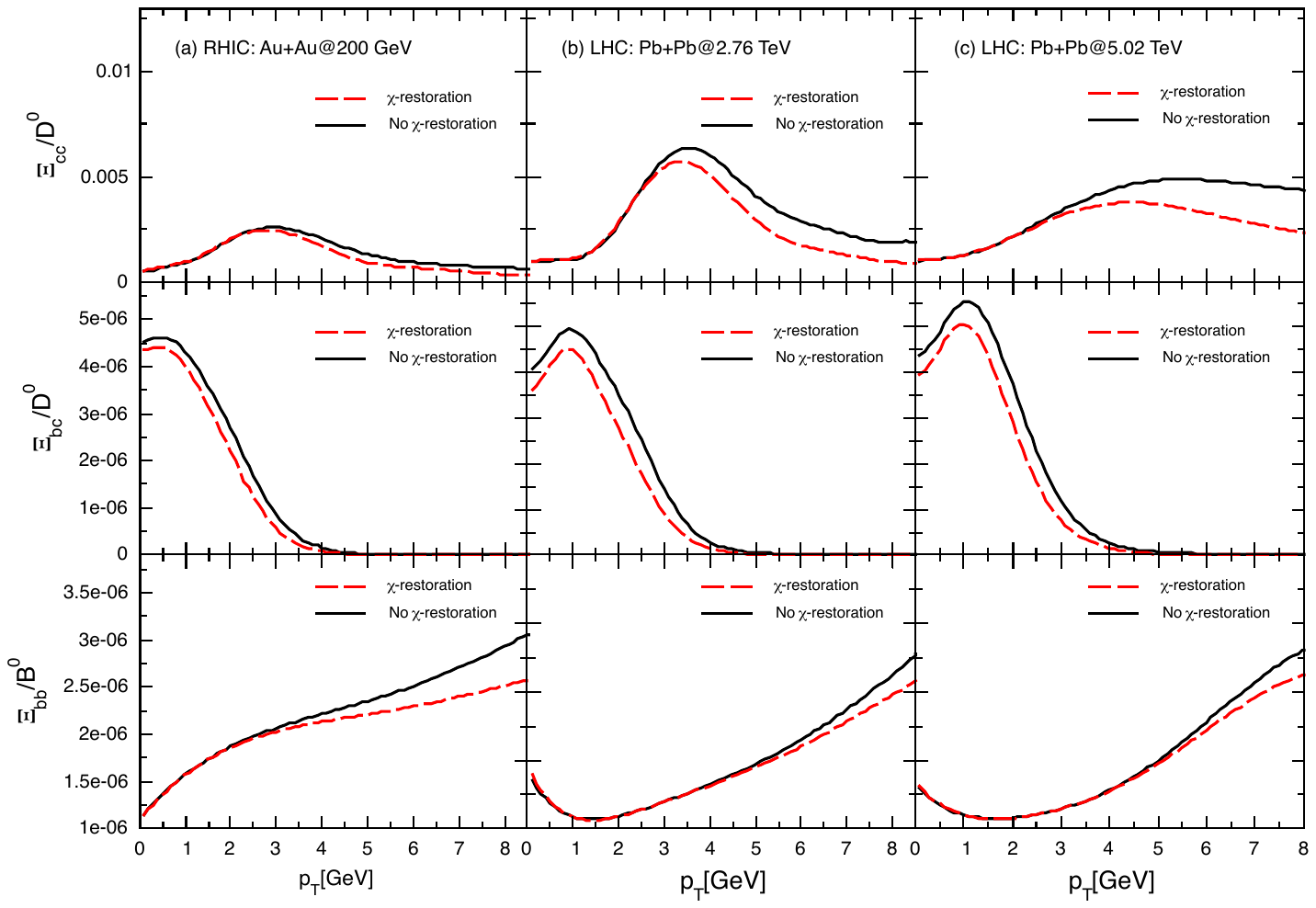}}
\\[-1.0true cm]
\caption{Same as Figure~\ref{ratio_charm} for $\Xi_{cc}/D^0$, $\Xi_{bc}/D^0$ and  $\Xi_{bb}/B^0$.}
\label{ratio_doublecascade}
\end{center}
\end{figure}

The transverse momentum spectra for baryon-to-meson ratios involving double cascade charm, $\Xi_{cc}/D^0$, bottom, $\Xi_{bb}/B^0$, and charm-bottom, $\Xi_{bc}/D^0$ baryons are shown in Fig.~\ref{ratio_doublecascade}.
Ratios involving double cascade baryons, such as $\Xi_{cc}$, $\Xi_{bc}$, and $\Xi_{bb}$, are significantly lower than the corresponding baryon-to-meson ratios involving single-charm and single-bottom $\Lambda_{Q}$, $\Sigma_{Q}$ and $\Xi_{Q}$ ($Q=c$ or $b$) baryons.
This suppression can be explained by the low probability of producing two heavy quarks in a single collision. The same trend as in $\Xi_{c}/D^0$ and $\Xi_b/B^0$ ratios is observed, with the effects of chiral symmetry restoration being less pronounced.
At intermediate transverse momentum, a peak is observed in the spectra, which is consistent with the typical behavior of hadrons produced via coalescence mechanisms. The position and height of the peak vary between the different baryon species $\Xi_{cc}$, $\Xi_{bc}$, and $\Xi_{bb}$. In particular, the heavier the baryon, the more suppressed the spectrum tends to be at a higher $p_T$, with the $\Xi_{bb}$ ratio showing the lowest values compared to the charm baryon sector.

\begin{figure}[t]
\begin{center}
\makebox[\textwidth][c]{\includegraphics[trim=1 1 1 90, clip,width=1.1\textwidth]{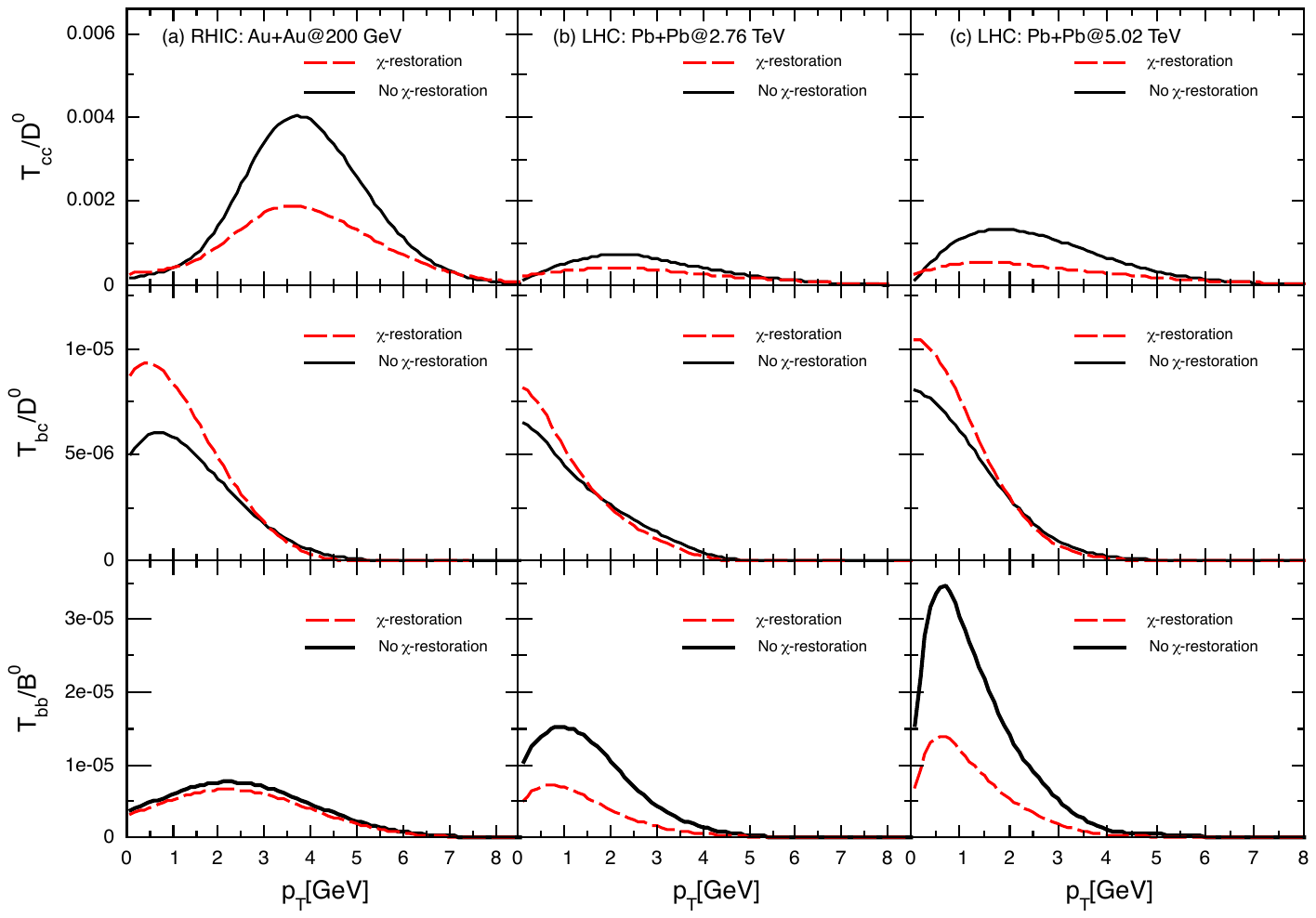}}
\\[-1.0true cm]
\caption{Same as Figure~\ref{ratio_charm} for $T_{cc}/D^0$, $T_{bc}/D^0$ and  $T_{bb}/B^0$.}
\label{ratio_tetraquark}
\end{center}
\end{figure}

Finally, in Fig.~\ref{ratio_tetraquark} we present results for the transverse momentum spectra of the tetraquark-to-meson ratios for the three $J^P=1^+$ tetraquarks: $T_{cc}$, $T_{bc}$ and $T_{bb}$. The ratios exhibit a strong dependence on $p_T$, peaking at low to intermediate $p_T$ values and decreasing quickly at higher $p_T$. Tetraquarks are relatively more abundant at low transverse momenta, while mesons become more dominant at higher $p_T$. Moreover, the effect of chiral symmetry restoration is more pronounced at low $p_T$. $\chi$-restoration effects tend to decrease the tetraquark-to-meson ratio for systems with two identical heavy quarks, while it enhances the production in the case of nonidentical heavy flavors. This is clearly due to the different components of the wave function of a double-heavy tetraquark with or without identical heavy flavors, see Ref.~\cite{Caramees:2018oue} for a detailed discussion. This also makes it clear that great care must be taken with simple single-Gaussian wave function hypotheses whose parameter is set to some physical observable, either the size of the hadron or the normalization of the Wigner function.

Before concluding, it should be noted that not all 
heavy quarks in the quark-gluon plasma participate in coalescence.In our framework, we focus on modeling the coalescence contribution to heavy quark hadronization. While a fully consistent calculation of the fragmentation component is beyond the scope of the present work, we can estimate the coalescence fraction for charm quarks by combining our model predictions with measured baryon-to-meson ratios in heavy-ion collisions~\cite{STAR2020,ALICE:2018hbc}, using pp baselines~\cite{ALICE:2018jhep} as reference. This approach implicitly assumes that the modification of the baryon-to-meson ratio in nucleus-nucleus collisions, compared to the pp baseline, is mainly driven by the presence of the quark-gluon plasma and the enhanced probability of quark recombination. The coalescence contribution is thus extracted by isolating the deviation from the vacuum fragmentation pattern observed experimentally.

Based on this procedure, we find that approximately 87\% of charm quarks hadronize via coalescence in central Au+Au collisions at RHIC energies ($\sqrt{s_{NN}} = 200$ GeV), decreasing to about 50\% at LHC energies ($\sqrt{s_{NN}} = 5.02$ TeV).This decrease reflects the interplay between the charm quark momentum distributions and the medium properties, with harder spectra and lower parton densities at higher energies favoring fragmentation over coalescence.
For bottom quarks, however, due to the current lack of systematic and conclusive experimental measurements of bottom hadron production in heavy-ion collisions, it is not possible to provide a reliable estimate of the coalescence fraction within our approach.

We note that our estimates for charm quark hadronization are consistent with previous theoretical and phenomenological studies~\cite{Oh:2009zj,Cho:2020,Greco:2018,Cao:2020}, which suggest that a significant fraction of charm quarks undergo coalescence at low transverse momentum ($p_T < 2$–$3$ GeV), with fragmentation gradually taking over at higher $p_T$.
This behavior can be understood by considering that, at low $p_T$, the charm quark kinematics are more compatible with the formation of bound states through phase-space proximity with light quarks, while at high $p_T$ the larger momentum mismatch reduces the coalescence probability and favors independent fragmentation.
These findings support the general understanding that coalescence dominates at low $p_T$ and fragmentation becomes more relevant at high $p_T$, in agreement with experimental observations~\cite{Greco:2018,Cao:2020} and theoretical expectations~\cite{Cho:2020,Greco:2018}.

\section{Conclusions and outlook}
\label{sec:6}

We have carried out a simultaneous study of the production of conventional and exotic open-flavor heavy hadrons in relativistic heavy-ion collisions in the quark coalescence model. We evaluated yields and transverse momentum distributions
using hadron wave functions obtained from a single realistic quark model.
For the first time, the yields and transverse momentum distributions for the expected axial-vector double heavy tetraquarks have been studied in a realistic model describing the low-energy hadron spectra.

Our results show that in a realistic model the yields of exotic hadronic matter, in particular the $T_{QQ}$ tetraquarks, are of the same order of magnitude as the corresponding double heavy baryons. This can be explained in terms of the internal structure of the $QQ\bar u\bar d$ tetraquarks. Thus, when the strategy for the detection of doubly heavy baryons is improved in current accelerators, an equivalent method for the detection of tetraquarks will be available through an increase in energy. This type of study is of relevance especially for the possible production of stable bottom tetraquarks. Having realistic predictions of the production of
such exotic hadrons compared to the known heavy hadrons is a basic tool when
designing experiments that can allow us to access these new states of matter.

Our results show that partial chiral symmetry restoration enhances the yields, being more pronounced for the lightest hadrons due to two different factors. Firstly, the increase of the hadron r.m.s. radius enhances the coalescence probability. Secondly, the number of light quarks also increases with increasing temperature, contributing to the enhancement of the yields. The impact of partial chiral symmetry on the yields of $T_{cc}$,  $T_{bc}$, and $T_{bb}$ is small. Although temperature significantly affects its spatial distribution, resulting in a decrease of the r.m.s. radius, this effect is practically offset by the increase of the number of light quarks, leading to negligible changes in the yields. It is also important to emphasize the increase of the yields due to the use of a fully relativistic expression for the light-quark distribution.

Finally, considering chiral symmetry restoration effects, 
we obtain $N_{\Lambda_c}/N_{D^0}=0.98$ at RHIC. For the LHC
at 2.76~TeV we obtain $N_{\Lambda_c}/N_{D^0}=0.94$ and at 5.02~TeV
we get $N_{\Lambda_c}/N_{D^0}=0.92$. For the bottom sector, we get
$N_{\Lambda_b}/N_{B^0}=0.75$ at RHIC,
$N_{\Lambda_b}/N_{B^0}=0.66$ at LHC at 2.76~GeV, and
$N_{\Lambda_b}/N_{B^0}=0.45$ at LHC at 5.02~GeV.
The presence of a strange quark mitigates the effects of chiral symmetry restoration on $\Xi_c$ baryons, leading to negligible changes in their production ratios compared to baryons containing only light quarks.

\section{Acknowledgements.}
This work has been partially funded by the Spanish Ministerio de Ciencia e 
Innovaci\'on (MICINN) and the European Regional Development Fund (ERDF), 
under contracts PID2019-105439GB-C22, PID2022-141910NB-I00, and RED2022-134204-E, by 
Junta de Castilla y Le\'on under contract SA091P24. Partial financial support is also acknowledged from Conselho Nacional de Desenvolvimento Cient\'{\i}fico e Tecnol\'ogico-CNPq, Grants Nos. 309262/2019-4 and 464898/2014-5, and Funda\c{c}\~ao de Amparo \`a Pesquisa do Estado de S\~ao Paulo-FAPESP, Grants Nos. 2018/25225-9 and 2016/01343-7.

\end{document}